\documentclass[10pt,a4paper,draft]{article}
\usepackage{amsfonts,amsmath,amssymb,cite,lscape}
\newcommand{\Real}{\mathop{\textrm{Re}}}

\begin{document}
\title{Second-order Stark effect and polarizability of a relativistic
two-dimensional hydrogen-like atom in the ground state}
\author{Rados{\l}aw Szmytkowski\footnote{Email:
radoslaw.szmytkowski@pg.edu.pl} \\*[3ex]
Atomic and Optical Physics Division, \\
Department of Atomic, Molecular and Optical Physics, \\
Faculty of Applied Physics and Mathematics,
Gda{\'n}sk University of Technology, \\
ul.\ Gabriela Narutowicza 11/12, 80--233 Gda{\'n}sk, Poland}
\date{}
\maketitle
\begin{abstract} 
The second-order Stark effect for a planar Dirac one-electron atom in
the ground state is analyzed within the framework of the
Rayleigh--Schr{\"o}dinger perturbation theory, with the use of the
Sturmian series expansion of the generalized Dirac--Coulomb Green
function. A closed-form analytical expression for the static dipole
polarizability of that system is found. The formula involves a
generalized hypergeometric function ${}_{3}F_{2}$ with the unit
argument. Numerical values of the polarizabilities for relativistic
planar hydrogenic atoms with atomic numbers $1\leqslant Z\leqslant68$
are provided in a tabular form. A simple formula for the
polarizability of a nonrelativistic two-dimensional hydrogenic atom,
reported previously by several other authors, is recovered from our
result in the nonrelativistic limit.
\vskip1ex
\noindent
\textbf{PACS 2010:} 03.65.Pm, 31.15.aj, 31.15.xp, 31.30.jc, 32.10.Dk,
32.60.+i
\end{abstract}
%
%%\newpage
%
\section{Introduction}
\label{I}
\setcounter{equation}{0}
For several decades, theoreticians have been investigating properties
of model planar quantum systems. Recent years have seen a growth of
interest in such studies, driven primarily by the rapid progress in
low-dimensional condensed matter physics and materials science. It is
not surprising that the system that has attracted much interest in
this context is the planar analogue of the one-electron atom. Its
exceptional charm is rooted in its physical simplicity, as well as in
the fact that the pertinent Schr{\"o}dinger, Klein--Gordon and Dirac
equations admit analytical solutions \cite{Yang91,Guo91}. In
consequence, a good deal of information about various properties of
that particular system has been gathered over the past years. However,
a somewhat astonishing asymmetry may be observed: whereas a number of
works have dealt with the planar hydrogenic atom subjected to the
action of a magnetic field (the reader will find a comprehensive
relevant bibliography in our recent works \cite{Szmy18a,Szmy18b}),
much less effort has been put to consider such an atom immersed in an
electric field
\cite{Lins74,Lede76,Tana87,Yang91,Fern92a,Fern92b,Adam92,Adam94,%
Haro98,Poku01,Ivan03} (cf.\ also Refs.\ \cite{Alli93,Pede07,Sael07,%
Pede16}). Further studies on the Stark effect for planar one-electron
atoms are thus desirable, and the present paper meets that demand.

In Refs.\ \cite{Lede76,Tana87,Yang91,Adam92,Fern92a,Fern92b,Alli93,%
Pede07,Sael07}, a simple analytical expression for the polarizability
of the two-dimensional hydrogen-like atom in the ground state has been
found (or may be inferred from akin results presented therein). A
common feature of all these works is that the atomic electron has been
described with the use of the Schr{\"o}dinger equation. In the present
paper, we shall derive an analytical formula for the polarizability of
that particular atomic system, but with the employment of the Dirac
equation rather than the Schr{\"o}dinger one. The calculations will be
carried out within the framework of the second-order
Rayleigh--Schr{\"o}dinger perturbation theory, with the use of the
Sturmian series expansion of the generalized radial Dirac--Coulomb
Green function. The resulting formula for the polarizability appears
to be much more complex that its nonrelativistic counterpart, and
involves an irreducible generalized hypergeometric function
${}_{3}F_{2}$ with the unit argument. In the nonrelativistic limit, we
recover the expression found in Refs.\
\cite{Lede76,Tana87,Yang91,Adam92,Fern92a,Fern92b,Alli93,Pede07,%
Sael07}.
%
%\newpage
%
\section{Problem formulation}
\label{II}
\setcounter{equation}{0}
We are concerned with a Dirac one-electron atom (or ion) with a
point-like, spinless and motionless nucleus of electric charge $+Ze$.
The atomic electron is constrained to move in a plane through the
nucleus. It is assumed that the interaction potential between the
electron and the nucleus is the one-over-distance Coulomb one. The
system is immersed in a static and uniform lateral electric field of
strength $\boldsymbol{F}$. It is posited that the electric field is
weak, in the sense that the electron is considered to stay bounded (in
other words, the probability of the occurrence of the field-ionization
process is negligibly small) and field-induced energy shifts are small
compared to the fine-structure splitting of the planar Dirac--Coulomb
energy levels. Under the assumptions specified above, the
time-independent electronic wave function is taken to be a solution to
the planar Dirac equation
\begin{subequations}
\begin{equation}
\left[-\mathrm{i}c\hbar\boldsymbol{\alpha}
\cdot\boldsymbol{\nabla}+\beta mc^{2}
-\frac{Ze^{2}}{(4\pi\epsilon_{0})r}+V^{(1)}(\boldsymbol{r})
-E\right]\Psi(\boldsymbol{r})=0
\qquad (\boldsymbol{r}\in\mathbb{R}^{2}),
\label{2.1a}
\end{equation}
subject to the standard constraint of single-valuedness and the
boundary conditions
\begin{equation}
\sqrt{r}\,\Psi(\boldsymbol{r})\stackrel{r\to0}{\longrightarrow}0,
\qquad
\sqrt{r}\,\Psi(\boldsymbol{r})\stackrel{r\to\infty}{\longrightarrow}0.
\label{2.1b}
\end{equation}
\label{2.1}%
\end{subequations}
In Eq.\ (\ref{2.1a}),
\begin{equation}
V^{(1)}(\boldsymbol{r})=e\boldsymbol{F}\cdot\boldsymbol{r}
\label{2.2}
\end{equation}
is the potential energy of the interaction between the electron and
the perturbing electric field $\boldsymbol{F}$. Henceforth, we shall
be assuming that the atomic plane is the $\{x,y\}$ one, with the
Cartesian unit vectors $\boldsymbol{n}_{x}$ and $\boldsymbol{n}_{y}$,
and that the field $\boldsymbol{F}$ is directed along
$\boldsymbol{n}_{x}$, i.e.,
\begin{equation}
\boldsymbol{F}=F\boldsymbol{n}_{x}.
\label{2.3}
\end{equation}
The two components of the Dirac vector matrix
\begin{equation}
\boldsymbol{\alpha}=\alpha_{1}\boldsymbol{n}_{x}
+\alpha_{2}\boldsymbol{n}_{y}
\label{2.4}
\end{equation}
and the matrix $\beta$ are taken to be
\begin{equation}
\alpha_{1}
=\left(
\begin{array}{cc}
0 & \sigma_{1} \\
\sigma_{1} & 0
\end{array}
\right),
\qquad
\alpha_{2}
=\left(
\begin{array}{cc}
0 & \sigma_{2} \\
\sigma_{2} & 0
\end{array}
\right),
\qquad
\beta
=\left(
\begin{array}{cc}
I & 0 \\
0 & -I
\end{array}
\right),
\label{2.5}
\end{equation}
respectively, where
\begin{equation}
\sigma_{1}
=\left(
\begin{array}{cc}
0 & 1 \\
1 & 0
\end{array}
\right),
\qquad
\sigma_{2}
=\left(
\begin{array}{cc}
0 & -\mathrm{i} \\
\mathrm{i} & 0
\end{array}
\right),
\label{2.6}
\end{equation}
while $I$ stands for the unit $2\times2$ matrix.

Since the field $\boldsymbol{F}$ has been assumed to be weak, we shall
treat the term $V^{(1)}(\boldsymbol{r})$ as a small perturbation of
the Dirac--Coulomb Hamiltonian. Proceeding within the framework of the
Rayleigh--Schr{\"o}dinger perturbation theory, $E$ and
$\Psi(\boldsymbol{r})$ may be sought in the forms of the series
\begin{subequations}
\begin{equation}
E=E^{(0)}+E^{(1)}+E^{(2)}+\cdots
\label{2.7a}
\end{equation}
and
\begin{equation}
\Psi(\boldsymbol{r})=\Psi^{(0)}(\boldsymbol{r})
+\Psi^{(1)}(\boldsymbol{r})+\Psi^{(2)}(\boldsymbol{r})+\cdots.
\label{2.7b}
\end{equation}
\label{2.7}%
\end{subequations}
The zeroth-order terms $E^{(0)}$ and $\Psi^{(0)}(\boldsymbol{r})$
appearing above are those solutions to the planar bound-state
Dirac--Coulomb energy-eigenvalue problem
\begin{subequations}
\begin{equation}
\left[-\mathrm{i}c\hbar\boldsymbol{\alpha}
\cdot\boldsymbol{\nabla}+\beta mc^{2}
-\frac{Ze^{2}}{(4\pi\epsilon_{0})r}-E^{(0)}\right]
\Psi^{(0)}(\boldsymbol{r})=0
\qquad (\boldsymbol{r}\in\mathbb{R}^{2}),
\label{2.8a}
\end{equation}
\begin{equation}
\sqrt{r}\,\Psi^{(0)}(\boldsymbol{r})
\stackrel{r\to0}{\longrightarrow}0,
\qquad
\sqrt{r}\,\Psi^{(0)}(\boldsymbol{r})
\stackrel{r\to\infty}{\longrightarrow}0,
\label{2.8b}
\end{equation}
\label{2.8}%
\end{subequations}
from which $E$ and $\Psi(\boldsymbol{r})$ evolve in effect of the
action of the electric field.

In this work, we shall carry out calculations of the energy
corrections $E^{(1)}$ and $E^{(2)}$ in the case when $E^{(0)}$ and
$\Psi^{(0)}(\boldsymbol{r})$ refer to the \emph{ground\/} state of the
isolated atom. The energy $E^{(0)}$ of that state is
\begin{equation}
E^{(0)}=2\gamma_{1/2}mc^{2},
\label{2.9}
\end{equation}
with
\begin{equation}
\gamma_{\kappa}=\sqrt{\kappa^{2}-(\alpha Z)^{2}},
\label{2.10}
\end{equation}
where $\alpha=e^{2}/(4\pi\epsilon_{0})c\hbar$ is the Sommerfeld
fine-structure constant, while the wave function
$\Psi^{(0)}(\boldsymbol{r})$ is
\begin{equation}
\Psi^{(0)}(\boldsymbol{r})
=a_{1/2}^{(0)}\Psi^{(0)}_{1/2}(\boldsymbol{r})
+a_{-1/2}^{(0)}\Psi^{(0)}_{-1/2}(\boldsymbol{r}),
\label{2.11}
\end{equation}
with the basis eigenfunctions chosen to be
\begin{equation}
\Psi_{m_{a}}^{(0)}(\boldsymbol{r})
=\frac{1}{\sqrt{r}}
\left(
\begin{array}{c}
P^{(0)}(r)\Phi_{-1/2,m_{a}}(\varphi) \\*[1ex]
\mathrm{i}Q^{(0)}(r)\Phi_{1/2,m_{a}}(\varphi)
\end{array}
\right)
\qquad \left(m_{a}=\pm{\textstyle\frac{1}{2}}\right)
\label{2.12}
\end{equation}
and with the mixing coefficients $a_{\pm1/2}^{(0)}$ constrained to
obey
\begin{equation}
|a_{1/2}^{(0)}|^{2}+|a_{-1/2}^{(0)}|^{2}=1.
\label{2.13}
\end{equation}
In Eq.\ (\ref{2.12}) and hereafter, $0\leqslant\varphi<2\pi$ is the
polar angle between the unit vector $\boldsymbol{n}_{x}$ and the
radius vector $\boldsymbol{r}$,
\begin{equation}
\Phi_{\kappa m_{\kappa}}(\varphi)
=\frac{1}{\sqrt{2\pi}}
\left(
\begin{array}{c}
\delta_{-\kappa,m_{\kappa}}\,
\mathrm{e}^{\mathrm{i}(m_{\kappa}-1/2)\varphi}
\\*[1ex]
\delta_{\kappa m_{\kappa}}\,
\mathrm{e}^{\mathrm{i}(m_{\kappa}+1/2)\varphi}
\end{array}
\right)
\qquad
\left(\textrm{${\textstyle\kappa=\pm\frac{1}{2},\pm\frac{3}{2},
\pm\frac{5}{2},\ldots}$; $m_{\kappa}=\pm\kappa$}\right)
\label{2.14}
\end{equation}
are the axial spinors introduced by Poszwa and Rutkowski \cite{Posz10}
and discussed more comprehensively in Ref.\ \cite[Appendix~A]{Szmy18b}
(notice that the quantum number $\kappa$ used in the present paper and
in Ref.\ \cite{Szmy18b} has the opposite sign in relation to the one
from Ref.\ \cite{Posz10}), while the ground-state radial functions
$P^{(0)}(r)$ and $Q^{(0)}(r)$ are
\begin{subequations}
\begin{equation}
P^{(0)}(r)=\sqrt{\frac{2Z(1+2\gamma_{1/2})}
{a_{0}\Gamma(2\gamma_{1/2}+1)}}
\left(\frac{4Zr}{a_{0}}\right)^{\gamma_{1/2}}
\mathrm{e}^{-2Zr/a_{0}}
\label{2.15a}
\end{equation}
and
\begin{equation}
Q^{(0)}(r)=\sqrt{\frac{2Z(1-2\gamma_{1/2})}
{a_{0}\Gamma(2\gamma_{1/2}+1)}}
\left(\frac{4Zr}{a_{0}}\right)^{\gamma_{1/2}}
\mathrm{e}^{-2Zr/a_{0}},
\label{2.15b}
\end{equation}
\label{2.15}%
\end{subequations}
with $a_{0}=(4\pi\epsilon_{0})\hbar^{2}/me^{2}$ being the Bohr radius.
To ensure that $\gamma_{1/2}$ is real and positive, we impose the
constraint
\begin{equation}
Z<\frac{1}{2}\alpha^{-1}.
\label{2.16}
\end{equation}

It may be verified that the axial spinors (\ref{2.14}) are orthonormal
in the sense of
\begin{equation}
\int_{0}^{2\pi}\mathrm{d}\varphi\:
\Phi_{\kappa m_{\kappa}}^{\dag}(\varphi)
\Phi_{\kappa'm_{\kappa}^{\prime}}(\varphi)
=\delta_{\kappa\kappa'}\delta_{m_{\kappa}m_{\kappa}^{\prime}}
\label{2.17}
\end{equation}
and that the radial functions (\ref{2.15}) are normalized to unity in
the sense of
\begin{equation}
\int_{0}^{\infty}\mathrm{d}r
\left\{[P^{(0)}(r)]^{2}+[Q^{(0)}(r)]^{2}\right\}=1.
\label{2.18}
\end{equation}
Consequently, it holds that
\begin{equation}
\int_{\mathbb{R}^{2}}\mathrm{d}^{2}\boldsymbol{r}\:
\Psi_{m_{a}}^{(0)\dag}(\boldsymbol{r})
\Psi_{m_{a}^{\prime}}^{(0)}(\boldsymbol{r})
=\delta_{m_{a}m_{a}^{\prime}}
\qquad
\left(m_{a},m_{a}^{\prime}=\pm{\textstyle\frac{1}{2}}\right)
\label{2.19}
\end{equation}
and, by virtue of the constraint (\ref{2.13}), the function
(\ref{2.11}) is normalized to unity in the sense of
\begin{equation}
\int_{\mathbb{R}^{2}}\mathrm{d}^{2}\boldsymbol{r}\:
\Psi^{(0)\dag}(\boldsymbol{r})\Psi^{(0)}(\boldsymbol{r})=1.
\label{2.20}
\end{equation}
%
%\newpage
%
\section{The first-order Stark effect}
\label{III}
\setcounter{equation}{0}
The first-order corrections $E^{(1)}$ and
$\Psi^{(1)}(\boldsymbol{r})$ appearing in Eqs.\ (\ref{2.7a}) and
(\ref{2.7b}) solve the inhomogeneous system
\begin{subequations}
\begin{equation}
\left[-\mathrm{i}c\hbar\boldsymbol{\alpha}
\cdot\boldsymbol{\nabla}+\beta mc^{2}
-\frac{Ze^{2}}{(4\pi\epsilon_{0})r}-E^{(0)}\right]
\Psi^{(1)}(\boldsymbol{r})
=-[V^{(1)}(\boldsymbol{r})-E^{(1)}]
\Psi^{(0)}(\boldsymbol{r}),
\label{3.1a}
\end{equation}
\begin{equation}
\sqrt{r}\,\Psi^{(1)}(\boldsymbol{r})
\stackrel{r\to0}{\longrightarrow}0,
\qquad
\sqrt{r}\,\Psi^{(1)}(\boldsymbol{r})
\stackrel{r\to\infty}{\longrightarrow}0,
\label{3.1b}
\end{equation}
\label{3.1}%
\end{subequations}
subject to the orthogonality constraint
\begin{equation}
\int_{\mathbb{R}^{2}}\mathrm{d}^{2}\boldsymbol{r}\:
\Psi_{m_{a}}^{(0)\dag}(\boldsymbol{r})
\Psi^{(1)}(\boldsymbol{r})=0
\qquad \left(m_{a}=\pm{\textstyle\frac{1}{2}}\right).
\label{3.2}
\end{equation}
Inserting Eq.\ (\ref{2.11}) into the right-hand side of Eq.\
(\ref{3.1a}), and then projecting the resulting equation from the left
onto the unperturbed eigenfunctions
$\Psi^{(0)}_{\pm1/2}(\boldsymbol{r})$, yields the algebraic system
\begin{equation}
\sum_{m_{a}^{\prime}=\pm1/2}[V_{m_{a}m_{a}^{\prime}}^{(1)}
-E^{(1)}\delta_{m_{a}m_{a}^{\prime}}]a_{m_{a}^{\prime}}^{(0)}=0
\qquad \left(m_{a}=\pm{\textstyle\frac{1}{2}}\right),
\label{3.3}
\end{equation}
with
\begin{equation}
V_{m_{a}m_{a}^{\prime}}^{(1)}
=\int_{\mathbb{R}^{2}}\mathrm{d}^{2}\boldsymbol{r}\:
\Psi_{m_{a}}^{(0)\dag}(\boldsymbol{r})V^{(1)}(\boldsymbol{r})
\Psi_{m_{a}^{\prime}}^{(0)}(\boldsymbol{r})
\qquad \left(m_{a},m_{a}^{\prime}=\pm{\textstyle\frac{1}{2}}\right).
\label{3.4}
\end{equation}
Since the perturbation operator (\ref{2.2}) may be written in the form
\begin{equation}
V^{(1)}(\boldsymbol{r})=eFr\cos\varphi,
\label{3.5}
\end{equation}
using Eq.\ (\ref{2.12}) and the integral identity
\begin{equation}
\int_{0}^{2\pi}\mathrm{d}\varphi\:\cos\varphi\,
\Phi_{\kappa m_{\kappa}}^{\dag}(\varphi)
\Phi_{\kappa'm_{\kappa}^{\prime}}(\varphi)
=\frac{1}{2}\delta_{m_{\kappa}/\kappa,m_{\kappa}^{\prime}/\kappa'}
(\delta_{\kappa,\kappa'+1}+\delta_{\kappa,\kappa'-1}),
\label{3.6}
\end{equation}
we infer that
\begin{equation}
V_{m_{a}m_{a}^{\prime}}^{(1)}=0
\qquad \left(m_{a},m_{a}^{\prime}=\pm{\textstyle\frac{1}{2}}\right).
\label{3.7}
\end{equation}
Consequently, the first-order contribution to the energy eigenvalue
$E$ vanishes:
\begin{equation}
E^{(1)}=0,
\label{3.8}
\end{equation}
and the mixing coefficients $a_{\pm1/2}^{(0)}$ remain undetermined at
this stage.

With the result (\ref{3.8}) taken into account, a formal solution to
the system (\ref{3.1}) is
\begin{equation}
\Psi^{(1)}(\boldsymbol{r})
=-\int_{\mathbb{R}^{2}}\mathrm{d}^{2}\boldsymbol{r}'\:
\hat{\mathcal{G}}^{(0)}(\boldsymbol{r},\boldsymbol{r}')
V^{(1)}(\boldsymbol{r}')\Psi^{(0)}(\boldsymbol{r}'),
\label{3.9}
\end{equation}
where
$\hat{\mathcal{G}}^{(0)}(\boldsymbol{r},\boldsymbol{r}')$ --- the
generalized planar Dirac--Coulomb Green function associated with the
unperturbed energy level (\ref{2.9}) --- is a solution to the
inhomogeneous system
\begin{subequations}
\begin{eqnarray}
&& \hspace*{-3em}
\left[-\mathrm{i}c\hbar\boldsymbol{\alpha}
\cdot\boldsymbol{\nabla}+\beta mc^{2}
-\frac{Ze^{2}}{(4\pi\epsilon_{0})r}-E^{(0)}\right]
\hat{\mathcal{G}}^{(0)}(\boldsymbol{r},\boldsymbol{r}')
=\delta^{(2)}(\boldsymbol{r}-\boldsymbol{r}')\mathcal{I}
-\sum_{m_{a}=\pm1/2}
\Psi_{m_{a}}^{(0)}(\boldsymbol{r})
\Psi_{m_{a}}^{(0)\dag}(\boldsymbol{r}')
\nonumber \\
&& \hspace*{28em} 
(\boldsymbol{r},\boldsymbol{r}'\in\mathbb{R}^{2}),
\label{3.10a}
\end{eqnarray}
\begin{equation}
\sqrt{r}\,\hat{\mathcal{G}}^{(0)}(\boldsymbol{r},\boldsymbol{r}')
\stackrel{r\to0}{\longrightarrow}0,
\qquad
\sqrt{r}\,\hat{\mathcal{G}}^{(0)}(\boldsymbol{r},\boldsymbol{r}')
\stackrel{r\to\infty}{\longrightarrow}0
\label{3.10b}
\end{equation}
\label{3.10}%
\end{subequations}
(here $\mathcal{I}$ is the unit $4\times4$ matrix), subject to the
disambiguating orthogonality constraint
\begin{equation}
\int_{\mathbb{R}^{2}}\mathrm{d}^{2}\boldsymbol{r}\:
\Psi_{m_{a}}^{(0)\dag}(\boldsymbol{r})
\hat{\mathcal{G}}^{(0)}(\boldsymbol{r},\boldsymbol{r}')=0
\qquad \left(m_{a}=\pm{\textstyle\frac{1}{2}}\right).
\label{3.11}
\end{equation}
The expression for $\Psi^{(1)}(\boldsymbol{r})$ given in Eq.\
(\ref{3.9}) will be used in the next section, where the second-order
Stark effect will be analyzed.
\section{The second-order Stark effect and the atomic
po\-la\-ri\-za\-bi\-li\-ty}
\label{IV}
\setcounter{equation}{0}
The second-order corrections $E^{(2)}$ and
$\Psi^{(2)}(\boldsymbol{r})$ are solutions to the inhomogeneous system
\begin{subequations}
\begin{eqnarray}
&& \left[-\mathrm{i}c\hbar\boldsymbol{\alpha}
\cdot\boldsymbol{\nabla}+\beta mc^{2}
-\frac{Ze^{2}}{(4\pi\epsilon_{0})r}-E^{(0)}\right]
\Psi^{(2)}(\boldsymbol{r})
=-[V^{(1)}(\boldsymbol{r})-E^{(1)}]\Psi^{(1)}(\boldsymbol{r})
+E^{(2)}\Psi^{(0)}(\boldsymbol{r}),
\nonumber \\
&&
\label{4.1a}
\end{eqnarray}
\begin{equation}
\sqrt{r}\,\Psi^{(2)}(\boldsymbol{r})
\stackrel{r\to0}{\longrightarrow}0,
\qquad
\sqrt{r}\,\Psi^{(2)}(\boldsymbol{r})
\stackrel{r\to\infty}{\longrightarrow}0,
\label{4.1b}
\end{equation}
\label{4.1}%
\end{subequations}
augmented with the orthogonality condition
\begin{equation}
\int_{\mathbb{R}^{2}}\mathrm{d}^{2}\boldsymbol{r}\:
\Psi_{m_{a}}^{(0)\dag}(\boldsymbol{r})
\Psi^{(2)}(\boldsymbol{r})=0
\qquad \left(m_{a}=\pm{\textstyle\frac{1}{2}}\right).
\label{4.2}
\end{equation}
Proceeding as in the preceding section, after making use of the
results (\ref{3.8}) and (\ref{3.9}), one arrives at the following
algebraic system for the thus far undetermined mixing coefficients
$a_{\pm1/2}^{(0)}$:
\begin{equation}
\sum_{m_{a}^{\prime}=\pm1/2}[V_{m_{a}m_{a}^{\prime}}^{(1,1)}
-E^{(2)}\delta_{m_{a}m_{a}^{\prime}}]a_{m_{a}^{\prime}}^{(0)}=0
\qquad \left(m_{a}=\pm{\textstyle\frac{1}{2}}\right),
\label{4.3}
\end{equation}
with
\begin{eqnarray}
&& V_{m_{a}m_{a}^{\prime}}^{(1,1)}
=-\int_{\mathbb{R}^{2}}\mathrm{d}^{2}\boldsymbol{r}\:
\int_{\mathbb{R}^{2}}\mathrm{d}^{2}\boldsymbol{r}'\:
\Psi_{m_{a}}^{(0)\dag}(\boldsymbol{r})
V^{(1)}(\boldsymbol{r})
\hat{\mathcal{G}}^{(0)}(\boldsymbol{r},\boldsymbol{r}')
V^{(1)}(\boldsymbol{r}')
\Psi_{m_{a}^{\prime}}^{(0)}(\boldsymbol{r}')
\nonumber \\
&& \hspace*{25em} 
\left(m_{a},m_{a}^{\prime}=\pm{\textstyle\frac{1}{2}}\right).
\label{4.4}
\end{eqnarray}
Plugging Eqs.\ (\ref{2.12}) and (\ref{3.5}), and also the following
multipole representation of
$\hat{\mathcal{G}}^{(0)}(\boldsymbol{r},\boldsymbol{r}')$:
\begin{eqnarray}
\hat{\mathcal{G}}^{(0)}(\boldsymbol{r},\boldsymbol{r}')
&=& \sum_{\kappa=-\infty-1/2}^{+\infty+1/2}
\sum_{m_{\kappa}=\pm\kappa}
\frac{1}{\sqrt{rr'}}
\nonumber \\
&& \hspace*{-5em}
\times\,\left(
\begin{array}{cc}
\hat{g}_{(++)\kappa}^{(0)}(r,r')
\Phi_{\kappa m_{\kappa}}(\varphi)
\Phi_{\kappa m_{\kappa}}^{\dag}(\varphi') &
-\mathrm{i}\hat{g}_{(+-)\kappa}^{(0)}(r,r')
\Phi_{\kappa m_{\kappa}}(\varphi)
\Phi_{-\kappa m_{\kappa}}^{\dag}(\varphi')
\\*[1ex]
\mathrm{i}\hat{g}_{(-+)\kappa}^{(0)}(r,r')
\Phi_{-\kappa m_{\kappa}}(\varphi)
\Phi_{\kappa m_{\kappa}}^{\dag}(\varphi') &
\hat{g}_{(--)\kappa}^{(0)}(r,r')
\Phi_{-\kappa m_{\kappa}}(\varphi)
\Phi_{-\kappa m_{\kappa}}^{\dag}(\varphi')
\end{array}
\right),
\nonumber \\
&&
\label{4.5}
\end{eqnarray}
into the right-hand side of Eq.\ (\ref{4.4}), and then carrying out
angular integrations with the aid of Eq.\ (\ref{3.6}), casts the
matrix element $V_{m_{a}m_{a}^{\prime}}^{(1,1)}$ into the form
\begin{eqnarray}
&& V_{m_{a}m_{a}^{\prime}}^{(1,1)}=-\delta_{m_{a}m_{a}^{\prime}}
\frac{1}{4}e^{2}F^{2}
\left[R_{1/2}^{(1,1)}\big(P^{(0)},Q^{(0)};P^{(0)},Q^{(0)}\big)
+R_{-3/2}^{(1,1)}\big(P^{(0)},Q^{(0)};P^{(0)},Q^{(0)}\big)\right]
\nonumber \\
&& \hspace*{27em} 
\left(m_{a},m_{a}^{\prime}=\pm{\textstyle\frac{1}{2}}\right),
\label{4.6}
\end{eqnarray}
with
\begin{eqnarray}
&& \hspace*{-5em}
R_{\kappa}^{(1,1)}\big(P^{(0)},Q^{(0)};P^{(0)},Q^{(0)}\big)
\nonumber \\
&=& \int_{0}^{\infty}\mathrm{d}r\int_{0}^{\infty}\mathrm{d}r'
\left(
\begin{array}{cc}
P^{(0)}(r) & Q^{(0)}(r)
\end{array}
\right)
r\hat{G}_{\kappa}^{(0)}(r,r')r'
\left(
\begin{array}{c}
P^{(0)}(r') \\
Q^{(0)}(r')
\end{array}
\right),
\label{4.7}
\end{eqnarray}
where
\begin{equation}
\hat{G}_{\kappa}^{(0)}(r,r')
=\left(
\begin{array}{cc}
\hat{g}_{(++)\kappa}^{(0)}(r,r') &
\hat{g}_{(+-)\kappa}^{(0)}(r,r') \\*[1ex]
\hat{g}_{(-+)\kappa}^{(0)}(r,r') &
\hat{g}_{(--)\kappa}^{(0)}(r,r')
\end{array}
\right)
\label{4.8}
\end{equation}
is the generalized radial Dirac--Coulomb Green function associated
with the unperturbed ground-state energy level (\ref{2.9}). It is seen
from Eq.\ (\ref{4.6}) that the matrix with the elements
$V_{m_{a}m_{a}^{\prime}}^{(1,1)}$ is a multiple of the $2\times2$ unit
matrix; in effect the secular equation for the algebraic system
(\ref{4.3}) has the double root
\begin{equation} 
E^{(2)}=-\frac{1}{4}e^{2}F^{2}
\left[R_{1/2}^{(1,1)}\big(P^{(0)},Q^{(0)};P^{(0)},Q^{(0)}\big)
+R_{-3/2}^{(1,1)}\big(P^{(0)},Q^{(0)};P^{(0)},Q^{(0)}\big)\right]
\label{4.9}
\end{equation}
and the mixing coefficients $a_{\pm1/2}^{(0)}$ again remain
undetermined.

To complete the task of calculation of the second-order energy
correction $E^{(2)}$, we have to evaluate the double radial integral
(\ref{4.7}) for $\kappa=1/2$ and for $\kappa=-3/2$. For that purpose,
we shall exploit the following series representation of the
generalized Green function $\hat{G}_{\kappa}^{(0)}(r,r')$:
\begin{eqnarray}
\hat{G}_{\kappa}^{(0)}(r,r')
&=& \sum_{n_{r}=-\infty}^{\infty}
\frac{1}{\mu_{n_{r}\kappa}^{(0)}-1}
\left(
\begin{array}{c}
S_{n_{r}\kappa}^{(0)}(r) \\*[1ex]
T_{n_{r}\kappa}^{(0)}(r)
\end{array}
\right)
\left(
\begin{array}{cc}
\mu_{n_{r}\kappa}^{(0)}S_{n_{r}\kappa}^{(0)}(r') & 
T_{n_{r}\kappa}^{(0)}(r')
\end{array}
\right)
\qquad (\kappa\neq-{\textstyle\frac{1}{2}}),
\nonumber \\
&&
\label{4.10}
\end{eqnarray}
with
\begin{equation}
\mu_{n_{r}\kappa}^{(0)}=\frac{|n_{r}|+\gamma_{\kappa}
+N_{n_{r}\kappa}}{\gamma_{1/2}+\frac{1}{2}},
\label{4.11}
\end{equation}
involving the pertinent radial Dirac--Coulomb Sturmian functions (cf.\
Ref.\ \cite[Sec.\ 3]{Szmy18b}) evaluated at the energy (\ref{2.9}):
\begin{subequations}
\begin{eqnarray}
S_{n_{r}\kappa}^{(0)}(r) 
&=& \sqrt{\frac{4\pi\epsilon_{0}}{e^{2}}
\frac{(1+2\gamma_{1/2})|n_{r}|!(|n_{r}|+2\gamma_{\kappa})}
{4ZN_{n_{r}\kappa}(N_{n_{r}\kappa}-\kappa)
\Gamma(|n_{r}|+2\gamma_{\kappa})}}\,
\left(\frac{4Zr}{a_{0}}\right)^{\gamma_{\kappa}}
\mathrm{e}^{-2Zr/a_{0}}
\nonumber \\
&& \times\left[L_{|n_{r}|-1}^{(2\gamma_{\kappa})}
\left(\frac{4Zr}{a_{0}}\right)
-\frac{N_{n_{r}\kappa}-\kappa}{|n_{r}|+2\gamma_{\kappa}}
L_{|n_{r}|}^{(2\gamma_{\kappa})}
\left(\frac{4Zr}{a_{0}}\right)\right]
\label{4.12a}
\end{eqnarray}
and
\begin{eqnarray}
T_{n_{r}\kappa}^{(0)}(r) 
&=& -\,\sqrt{\frac{4\pi\epsilon_{0}}{e^{2}}
\frac{(1-2\gamma_{1/2})|n_{r}|!(|n_{r}|+2\gamma_{\kappa})}
{4ZN_{n_{r}\kappa}(N_{n_{r}\kappa}-\kappa)
\Gamma(|n_{r}|+2\gamma_{\kappa})}}\,
\left(\frac{4Zr}{a_{0}}\right)^{\gamma_{\kappa}}
\mathrm{e}^{-2Zr/a_{0}}
\nonumber \\
&& \times\left[L_{|n_{r}|-1}^{(2\gamma_{\kappa})}
\left(\frac{4Zr}{a_{0}}\right)
+\frac{N_{n_{r}\kappa}-\kappa}{|n_{r}|+2\gamma_{\kappa}}
L_{|n_{r}|}^{(2\gamma_{\kappa})}
\left(\frac{4Zr}{a_{0}}\right)\right].
\label{4.12b}
\end{eqnarray}
\label{4.12}%
\end{subequations}
Here $L_{n}^{(\alpha)}(\rho)$ is the generalized Laguerre polynomial
\cite[Sec.\ 5.5]{Magn66} [we define $L_{-1}^{(\alpha)}(\rho)\equiv0$],
and
\begin{equation}
N_{n_{r}\kappa}=\pm\sqrt{n_{r}^{2}+2|n_{r}|\gamma_{\kappa}
+\kappa^{2}},
\label{4.13}
\end{equation}
where one chooses the positive sign for $n_{r}>0$ and the negative
sign for $n_{r}<0$; if $n_{r}=0$, then the positive sign is to be
chosen for $\kappa\leqslant-\frac{1}{2}$ and the negative one for
$\kappa\geqslant\frac{1}{2}$, i.e., it holds that
$N_{0\kappa}=-\kappa$. The functions (\ref{4.12}) and the expansion
(\ref{4.10}) may be constructed proceeding along the route analogous
to the one taken by us in Ref.\ \cite{Szmy97} for the
three-dimensional Dirac--Coulomb problem.

Insertion of the expansion (\ref{4.10}) into Eq.\ (\ref{4.7})
transforms the latter into
\begin{eqnarray}
R_{\kappa}^{(1,1)}\big(P^{(0)},Q^{(0)};P^{(0)},Q^{(0)}\big)
&=& \sum_{n_{r}=-\infty}^{\infty}\frac{1}{\mu_{n_{r}\kappa}^{(0)}-1}
\int_{0}^{\infty}\mathrm{d}r\:r
\left[P^{(0)}(r)S_{n_{r}\kappa}^{(0)}(r)
+Q^{(0)}(r)T_{n_{r}\kappa}^{(0)}(r)\right]
\nonumber \\
&& \quad \times\int_{0}^{\infty}\mathrm{d}r'\:r'
\left[\mu_{n_{r}\kappa}^{(0)}P^{(0)}(r')S_{n_{r}\kappa}^{(0)}(r')
+Q^{(0)}(r')T_{n_{r}\kappa}^{(0)}(r')\right].
\nonumber \\
&&
\label{4.14}
\end{eqnarray}
The two radial integrals which enter the summand may be evaluated with
the use of Eqs.\ (\ref{2.15}), (\ref{4.11}) and (\ref{4.12}), together
with the known integration formula
\begin{equation}
\int_{0}^{\infty}\mathrm{d}\rho\:\rho^{\gamma}\mathrm{e}^{-\rho}
L_{n}^{(\alpha)}(\rho)=\frac{\Gamma(\gamma+1)\Gamma(n+\alpha-\gamma)}
{n!\Gamma(\alpha-\gamma)}
\qquad (\Real\gamma>-1).
\label{4.15}
\end{equation}
The results are
\begin{subequations}
\begin{eqnarray}
&& \hspace*{-5em}
\int_{0}^{\infty}\mathrm{d}r\:r
\left[P^{(0)}(r)S_{n_{r}\kappa}^{(0)}(r)
+Q^{(0)}(r)T_{n_{r}\kappa}^{(0)}(r)\right]
\nonumber \\
&=& -\frac{\sqrt{4\pi\epsilon_{0}}\,a_{0}^{3/2}}{e}
\frac{(N_{n_{r}\kappa}-\kappa)
[(|n_{r}|+\gamma_{\kappa}-\gamma_{1/2}-2)
-2\gamma_{1/2}(N_{n_{r}\kappa}+\kappa)]}
{8Z^{2}\sqrt{2|n_{r}|!N_{n_{r}\kappa}(N_{n_{r}\kappa}-\kappa)
\Gamma(2\gamma_{1/2}+1)\Gamma(|n_{r}|+2\gamma_{\kappa}+1)}}
\nonumber \\
&& \times\,\frac{\Gamma(\gamma_{\kappa}+\gamma_{1/2}+2)
\Gamma(|n_{r}|+\gamma_{\kappa}-\gamma_{1/2}-2)}
{\Gamma(\gamma_{\kappa}-\gamma_{1/2}-1)}
\label{4.16a}
\end{eqnarray}
and
\begin{eqnarray}
&& \hspace*{-5em}
\int_{0}^{\infty}\mathrm{d}r\:
r\left[\mu_{n_{r}\kappa}^{(0)}P^{(0)}(r)S_{n_{r}\kappa}^{(0)}(r)
+Q^{(0)}(r)T_{n_{r}\kappa}^{(0)}(r)\right]
\nonumber \\
&=& -\frac{\sqrt{4\pi\epsilon_{0}}\,a_{0}^{3/2}}{e}
\frac{\big(\mu_{n_{r}\kappa}^{(0)}-1\big)(N_{n_{r}\kappa}-\kappa)}
{16Z^{2}\sqrt{2|n_{r}|!N_{n_{r}\kappa}(N_{n_{r}\kappa}-\kappa)
\Gamma(2\gamma_{1/2}+1)\Gamma(|n_{r}|+2\gamma_{\kappa}+1)}}
\nonumber \\
&& \times\,\frac{\Gamma(\gamma_{\kappa}+\gamma_{1/2}+2)
\Gamma(|n_{r}|+\gamma_{\kappa}-\gamma_{1/2}-2)}
{\Gamma(\gamma_{\kappa}-\gamma_{1/2}-1)}
\nonumber \\
&& \times\bigg\{2\gamma_{1/2}(|n_{r}|+\gamma_{\kappa}-\gamma_{1/2}-2)
-(N_{n_{r}\kappa}+\kappa)
\nonumber \\
&& \quad +\,\frac{N_{n_{r}\kappa}+\frac{1}{2}}
{|n_{r}|+\gamma_{\kappa}-\gamma_{1/2}}
[(|n_{r}|+\gamma_{\kappa}-\gamma_{1/2}-2)
-2\gamma_{1/2}(N_{n_{r}\kappa}+\kappa)]\bigg\}.
\label{4.16b}
\end{eqnarray}
\label{4.16}%
\end{subequations}
Plugging Eqs.\ (\ref{4.16}) and (\ref{4.11}) into the right-hand side
of Eq.\ (\ref{4.14}), collecting then the terms corresponding to
$n_{r}$ and $-n_{r}$, after some algebra one arrives at the following
representation of the double integral (\ref{4.7}):
\begin{eqnarray}
&& \hspace*{-5em}
R_{\kappa}^{(1,1)}\big(P^{(0)},Q^{(0)};P^{(0)},Q^{(0)}\big)
\nonumber \\
&=& \frac{(4\pi\epsilon_{0})a_{0}^{3}}{e^{2}}
\frac{\Gamma^{2}(\gamma_{\kappa}+\gamma_{1/2}+2)}
{64Z^{4}\Gamma(2\gamma_{1/2}+1)\Gamma(2\gamma_{\kappa}+1)}
\nonumber \\
&& \times\bigg\{\frac{\gamma_{1/2}[(2\kappa+1)\gamma_{1/2}+4]}
{\gamma_{\kappa}-\gamma_{1/2}+1}\,
{}_{3}F_{2}
\left(
\begin{array}{c}
\gamma_{\kappa}-\gamma_{1/2}-1,\:
\gamma_{\kappa}-\gamma_{1/2}-1,\:
\gamma_{\kappa}-\gamma_{1/2}+1 \\
\gamma_{\kappa}-\gamma_{1/2}+2,\:
2\gamma_{\kappa}+1
\end{array}
;1
\right)
\nonumber \\
&& \quad
-\,\frac{\gamma_{\kappa}+\gamma_{1/2}}{2\kappa+1}\,
{}_{3}F_{2}
\left(
\begin{array}{c}
\gamma_{\kappa}-\gamma_{1/2}-1,\:
\gamma_{\kappa}-\gamma_{1/2}-1,\:
\gamma_{\kappa}-\gamma_{1/2} \\
\gamma_{\kappa}-\gamma_{1/2}+1,\:
2\gamma_{\kappa}+1
\end{array}
;1
\right)
\bigg\}
\nonumber \\
&& \hspace*{25em} 
\left(\textrm{$\kappa={\textstyle\frac{1}{2}}$ 
or $\kappa=-\textstyle{\frac{3}{2}}$}\right).
\label{4.17}
\end{eqnarray}
Here and hereafter, ${}_{3}F_{2}(\cdots)$ denotes the generalized
hypergeometric function
\begin{equation}
{}_{3}F_{2}
\left(
\begin{array}{c}
a_{1},\:
a_{2},\:
a_{3} \\
b_{1},\:
b_{2}
\end{array}
;z
\right)
=\frac{\Gamma(b_{1})\Gamma(b_{2})}
{\Gamma(a_{1})\Gamma(a_{2})\Gamma(a_{3})}\sum_{k=0}^{\infty}
\frac{\Gamma(a_{1}+k)\Gamma(a_{2}+k)\Gamma(a_{3}+k)}
{\Gamma(b_{1}+k)\Gamma(b_{2}+k)}\frac{z^{k}}{k!}.
\label{4.18}
\end{equation}
Application of the identity
\begin{eqnarray}
{}_{3}F_{2}
\left(
\begin{array}{c}
a_{1},\:
a_{2},\:
a_{3} \\
a_{3}+1,\:
b
\end{array}
;1
\right)
&=& \frac{\Gamma(b)\Gamma(b-a_{1}-a_{2}+1)}
{(b-a_{3}-1)\Gamma(b-a_{1})\Gamma(b-a_{2})}
\nonumber \\
&& -\,\frac{(a_{1}-a_{3}-1)(a_{2}-a_{3}-1)}{(a_{3}+1)(b-a_{3}-1)}
\,{}_{3}F_{2}
\left(
\begin{array}{c}
a_{1},\:
a_{2},\:
a_{3}+1 \\
a_{3}+2,\:
b
\end{array}
;1
\right)
\nonumber \\
&& \hspace*{10em} [\Real(b-a_{1}-a_{2})>-1]
\label{4.19}
\end{eqnarray}
brings the expression (\ref{4.17}) to the final general form
\begin{eqnarray}
&& \hspace*{-5em}
R_{\kappa}^{(1,1)}\big(P^{(0)},Q^{(0)};P^{(0)},Q^{(0)}\big)
\nonumber \\
&=& -\frac{(4\pi\epsilon_{0})a_{0}^{3}}{e^{2}}
\frac{(\gamma_{1/2}+1)(2\gamma_{1/2}+1)
(2\gamma_{1/2}+3)}{32Z^{4}(2\kappa+1)}
\nonumber \\
&& \times\bigg\{1-\frac{[(2\kappa+1)\gamma_{1/2}+2]^{2}
\Gamma^{2}(\gamma_{\kappa}+\gamma_{1/2}+2)}
{(\gamma_{\kappa}-\gamma_{1/2}+1)
\Gamma(2\gamma_{1/2}+4)\Gamma(2\gamma_{\kappa}+1)}
\nonumber \\
&& \quad
\times\,{}_{3}F_{2}
\left(
\begin{array}{c}
\gamma_{\kappa}-\gamma_{1/2}-1,\:
\gamma_{\kappa}-\gamma_{1/2}-1,\:
\gamma_{\kappa}-\gamma_{1/2}+1 \\
\gamma_{\kappa}-\gamma_{1/2}+2,\:
2\gamma_{\kappa}+1
\end{array}
;1
\right)
\bigg\}
\nonumber \\
&& \hspace*{20em} \left(\textrm{$\kappa={\textstyle\frac{1}{2}}$ 
or $\kappa=-\textstyle{\frac{3}{2}}$}\right).
\label{4.20}
\end{eqnarray}
For $\kappa=1/2$, the hypergeometric series in Eq.\ (\ref{4.20}) is a
truncating one and may be expressed in terms of elementary algebraic
functions. In that case one has
\begin{subequations}
\begin{equation}
R_{1/2}^{(1,1)}\big(P^{(0)},Q^{(0)};P^{(0)},Q^{(0)}\big)
=\frac{(4\pi\epsilon_{0})a_{0}^{3}}{e^{2}}
\frac{\gamma_{1/2}(\gamma_{1/2}+1)(2\gamma_{1/2}+1)
(4\gamma_{1/2}+5)}{64Z^{4}},
\label{4.21a}
\end{equation}
while for $\kappa=-3/2$ Eq.\ (\ref{4.20}) becomes
\begin{eqnarray}
&& \hspace*{-5em}
R_{-3/2}^{(1,1)}\big(P^{(0)},Q^{(0)};P^{(0)},Q^{(0)}\big)
\nonumber \\
&=& \frac{(4\pi\epsilon_{0})a_{0}^{3}}{e^{2}}
\frac{(\gamma_{1/2}+1)(2\gamma_{1/2}+1)
(2\gamma_{1/2}+3)}{64Z^{4}}
\nonumber \\
&& \times\bigg\{1-\frac{4(\gamma_{1/2}-1)^{2}
\Gamma^{2}(\gamma_{3/2}+\gamma_{1/2}+2)}
{(\gamma_{3/2}-\gamma_{1/2}+1)
\Gamma(2\gamma_{1/2}+4)\Gamma(2\gamma_{3/2}+1)}
\nonumber \\
&& \quad
\times\,{}_{3}F_{2}
\left(
\begin{array}{c}
\gamma_{3/2}-\gamma_{1/2}-1,\:
\gamma_{3/2}-\gamma_{1/2}-1,\:
\gamma_{3/2}-\gamma_{1/2}+1 \\
\gamma_{3/2}-\gamma_{1/2}+2,\:
2\gamma_{3/2}+1
\end{array}
;1
\right)
\bigg\}.
\label{4.21b}
\end{eqnarray}
\label{4.21}%
\end{subequations}
Hence, after Eqs.\ (\ref{4.21a}) and (\ref{4.21b}) are plugged into
Eq.\ (\ref{4.9}), the second-order correction to energy is found to be
\begin{eqnarray}
E^{(2)} &=& -\frac{(\gamma_{1/2}+1)^{2}(2\gamma_{1/2}+1)
(4\gamma_{1/2}+3)}{256Z^{4}}
\nonumber \\
&& \times\bigg\{1-\frac{4(\gamma_{1/2}-1)^{2}
\Gamma^{2}(\gamma_{3/2}+\gamma_{1/2}+2)}
{(\gamma_{1/2}+1)(4\gamma_{1/2}+3)(\gamma_{3/2}-\gamma_{1/2}+1)
\Gamma(2\gamma_{1/2}+3)\Gamma(2\gamma_{3/2}+1)}
\nonumber \\
&& \quad
\times\,{}_{3}F_{2}
\left(
\begin{array}{c}
\gamma_{3/2}-\gamma_{1/2}-1,\:
\gamma_{3/2}-\gamma_{1/2}-1,\:
\gamma_{3/2}-\gamma_{1/2}+1 \\
\gamma_{3/2}-\gamma_{1/2}+2,\:
2\gamma_{3/2}+1
\end{array}
;1
\right)
\bigg\}
\frac{F^{2}}{F_{0}^{2}}
\frac{e^{2}}{(4\pi\epsilon_{0})a_{0}},
\nonumber \\
&&
\label{4.22}
\end{eqnarray}
where
\begin{equation}
F_{0}=\frac{e}{(4\pi\epsilon_{0})a_{0}^{2}}
\simeq5.14\times10^{11}\,\mathrm{V/m}
\label{4.23}
\end{equation}
is the atomic unit of electric field.

The relationship between the second-order energy correction and the
strength of the perturbing electric field may be written in the form
\begin{equation}
E^{(2)}=-\frac{1}{2}(4\pi\epsilon_{0})\alpha_{1}F^{2},
\label{4.24}
\end{equation}
which defines the polarizability $\alpha_{1}$ of the system under
study. Comparison of Eqs.\ (\ref{4.24}) and (\ref{4.22}) yields the
following closed-form expression for the polarizability of the planar
Dirac one-electron atom in the ground state:
\begin{eqnarray}
\alpha_{1} &=& \frac{a_{0}^{3}}{Z^{4}}
\frac{(\gamma_{1/2}+1)^{2}(2\gamma_{1/2}+1)(4\gamma_{1/2}+3)}{128}
\nonumber \\
&& \times\bigg\{1-\frac{4(\gamma_{1/2}-1)^{2}
\Gamma^{2}(\gamma_{3/2}+\gamma_{1/2}+2)}
{(\gamma_{1/2}+1)(4\gamma_{1/2}+3)(\gamma_{3/2}-\gamma_{1/2}+1)
\Gamma(2\gamma_{1/2}+3)\Gamma(2\gamma_{3/2}+1)}
\nonumber \\
&& \quad
\times\,{}_{3}F_{2}
\left(
\begin{array}{c}
\gamma_{3/2}-\gamma_{1/2}-1,\:
\gamma_{3/2}-\gamma_{1/2}-1,\:
\gamma_{3/2}-\gamma_{1/2}+1 \\
\gamma_{3/2}-\gamma_{1/2}+2,\:
2\gamma_{3/2}+1
\end{array}
;1
\right)
\bigg\}.
\label{4.25}
\end{eqnarray}

Numerical results for the scaled polarizabilities $Z^{4}\alpha_{1}(Z)$
for planar hydrogenic atoms with \mbox{$1\leqslant Z\leqslant68$},
computed from Eq.\ (\ref{4.25}), are listed in Table \ref{T.1}. The
value of the inverse of the fine-structure constant used in
calculations has been $\alpha^{-1}=137.035\,999\,139$ (from CODATA
2014 \cite{Mohr16}). The data are displayed in the form which also
shows an estimated error in last two digits of each entry, resulting
from the declared one-standard-deviation uncertainty (equal to 31) in
the last two digits of the value of $\alpha^{-1}$ given above.
\begin{center}
[Place for Table \ref{T.1}]
\end{center}

It remains to investigate the expression in Eq.\ (\ref{4.25})  in the
quasi-relativistic limit $\alpha Z\ll1$. Using the approximations
\begin{equation}
\gamma_{\kappa}\simeq|\kappa|-\frac{(\alpha Z)^{2}}{2|\kappa|}
+\mathrm{O}\left((\alpha Z)^{4}\right),
\label{4.26}
\end{equation}
\begin{equation}
\Gamma(a\gamma_{\kappa}+a'\gamma_{\kappa'}+b)
\simeq\Gamma(a|\kappa|+a'|\kappa'|+b)
\left[1-\frac{(\alpha Z)^{2}}{2}
\left(\frac{a}{|\kappa|}+\frac{a'}{|\kappa'|}\right)
\psi(a|\kappa|+a'|\kappa'|+b)\right]
+\mathrm{O}\left((\alpha Z)^{4}\right),
\label{4.27}
\end{equation}
where $\psi(z)$ is the digamma function defined as
\begin{equation}
\psi(z)=\frac{1}{\Gamma(z)}\frac{\mathrm{d}\Gamma(z)}{\mathrm{d}z},
\label{4.28}
\end{equation}
and
\begin{equation}
{}_{3}F_{2}
\left(
\begin{array}{c}
\gamma_{3/2}-\gamma_{1/2}-1,\:
\gamma_{3/2}-\gamma_{1/2}-1,\:
\gamma_{3/2}-\gamma_{1/2}+1 \\
\gamma_{3/2}-\gamma_{1/2}+2,\:
2\gamma_{3/2}+1
\end{array}
;1
\right)\simeq1+\mathrm{O}\left((\alpha Z)^{4}\right),
\label{4.29}
\end{equation}
after straightforward but somewhat lengthy algebraic manipulations one
arrives at the following quasi-relativistic estimate of the
polarizability:
\begin{equation}
\alpha_{1}\simeq\alpha_{1}^{\mathrm{NR}}
\left[1-\frac{7}{2}(\alpha Z)^{2}\right]
+\mathrm{O}\left((\alpha Z)^{4}\right).
\label{4.30}
\end{equation}
Here
\begin{equation}
\alpha_{1}^{\mathrm{NR}}=\frac{21}{128}\frac{a_{0}^{3}}{Z^{4}}
\label{4.31}
\end{equation}
is the polarizability of the nonrelativistic planar one-electron atom
in the ground state. The expression in Eq.\ (\ref{4.31}) is identical
with the one derived independently, from purely nonrelativistic
considerations, by several other authors
\cite{Lede76,Tana87,Yang91,Adam92,Fern92a,Fern92b,Alli93,Pede07,%
Sael07}.

It is instructive to compare the formulas in Eqs.\ (\ref{4.25}),
(\ref{4.30}) and (\ref{4.31}) with their counterparts for the
three-dimensional one-electron atom, which are provided in Appendix
\ref{A}.
\appendix
\section{Appendix: Polarizability of a relativistic three-dimensional
hydrogenic atom in the ground state}
\label{A}
\setcounter{equation}{0}
The polarizability of the three-dimensional Dirac one-electron atom in
the ground state is
\begin{eqnarray}
\alpha_{1} &=& \frac{a_{0}^{3}}{Z^{4}}
\frac{(\gamma_{1}+1)(2\gamma_{1}+1)
(4\gamma_{1}^{2}+13\gamma_{1}+12)}{36}
\nonumber \\
&& \times\bigg\{1-\frac{2(\gamma_{1}-2)^{2}
\Gamma^{2}(\gamma_{2}+\gamma_{1}+2)}
{(\gamma_{1}+1)(4\gamma_{1}^{2}+13\gamma_{1}+12)
(\gamma_{2}-\gamma_{1}+1)\Gamma(2\gamma_{1}+2)\Gamma(2\gamma_{2}+1)}
\nonumber \\
&& \quad
\times\,{}_{3}F_{2}
\left(
\begin{array}{c}
\gamma_{2}-\gamma_{1}-1,\:
\gamma_{2}-\gamma_{1}-1,\:
\gamma_{2}-\gamma_{1}+1 \\
\gamma_{2}-\gamma_{1}+2,\:
2\gamma_{2}+1
\end{array}
;1
\right)
\bigg\}
\label{A.1}
\end{eqnarray}
(cf.\ Ref.\ \cite[Eq.\ (3.24)]{Szmy04}, Ref.\ \cite[Eq.\ (16)]{Yakh03}
and Ref.\ \cite[Eq.\ (3.42)]{Szmy16}), with $\gamma_{\kappa}$ defined
as in Eq.\ (\ref{2.10}). The quasi-relativistic limit of the
expression displayed in Eq.\ (\ref{A.1}) is
\begin{equation}
\alpha_{1}\simeq\alpha_{1}^{\mathrm{NR}}
\left[1-\frac{28}{27}(\alpha Z)^{2}\right]
+\mathrm{O}\left((\alpha Z)^{4}\right),
\label{A.2}
\end{equation}
where
\begin{equation}
\alpha_{1}^{\mathrm{NR}}=\frac{9}{2}\frac{a_{0}^{3}}{Z^{4}}
\label{A.3}
\end{equation}
is the polarizability of the nonrelativistic three-dimensional
hydrogenic atom in the ground state.
%
%\newpage
%

%
\begin{landscape}
\begin{table}[h!]
\caption{$Z^{4}$-scaled polarizabilities for planar Dirac one-electron
atoms in the ground state, computed from the analytical formula in
Eq.\ (\ref{4.25}). The number in parentheses following each entry is
an uncertainty in its last two digits, and stems from the
one-standard-deviation uncertainty (equal to 31) in the last two
digits of the value of the inverse of the fine-structure constant
$\alpha^{-1}=137.035\:999\:139$ (from CODATA 2014) used in
calculations. The nonrelativistic limit of $Z^{4}\alpha_{1}(Z)$ is
independent of $Z$ and equals
$Z^{4}\alpha_{1}^{\textrm{NR}}(Z)=0.164\,062\,5$~$a_{0}^{3}$ [cf.\
Eq.\ (\ref{4.31})]. \protect\\*[0ex]\protect}
\label{T.1}
\begin{center}
\begin{tabular}{rclccrclccrcl}
\hline \\*[-0.5ex]
\multicolumn{1}{c}{$Z$} &&
\multicolumn{1}{c}{$Z^{4}\alpha_{1}(Z)\:\:\:(a_{0}^{3})$} &&&
\multicolumn{1}{c}{$Z$} && 
\multicolumn{1}{c}{$Z^{4}\alpha_{1}(Z)\:\:\:(a_{0}^{3})$} &&& 
\multicolumn{1}{c}{$Z$} && 
\multicolumn{1}{c}{$Z^{4}\alpha_{1}(Z)\:\:\:(a_{0}^{3})$} \\*[1ex]
\hline \\*[-0.5ex]
1  && $0.164\,031\,922\,357\,129\:(14)$ &&&
24 && $0.146\,540\,774\,615\,7\:(79)$ &&&
47 && $0.097\,649\,864\,741\:(30)$ \\
2  && $0.163\,940\,192\,827\,883\:(55)$ &&&
25 && $0.145\,058\,299\,164\,6\:(86)$ &&&
48 && $0.094\,824\,238\,984\:(31)$ \\
3  && $0.163\,787\,321\,605\,63\:(12)$ &&&
26 && $0.143\,516\,593\,431\,0\:(92)$ &&&
49 && $0.091\,938\,523\,941\:(32)$ \\
4  && $0.163\,573\,325\,663\,45\:(22)$ &&&
27 && $0.141\,915\,791\,499\:(10)$ &&&
50 && $0.088\,991\,948\,689\:(34)$ \\
5  && $0.163\,298\,228\,730\,23\:(35)$ &&&
28 && $0.140\,256\,027\,891\:(11)$ &&&
51 && $0.085\,983\,547\,673\:(35)$ \\
6  && $0.162\,962\,061\,256\,62\:(50)$ &&&
29 && $0.138\,537\,436\,674\:(11)$ &&&
52 && $0.082\,912\,115\,141\:(37)$ \\
7  && $0.162\,564\,860\,370\,78\:(68)$ &&&
30 && $0.136\,760\,150\,441\:(12)$ &&&
53 && $0.079\,776\,146\,327\:(38)$ \\
8  && $0.162\,106\,669\,823\,04\:(88)$ &&&
31 && $0.134\,924\,299\,151\:(13)$ &&&
54 && $0.076\,573\,760\,514\:(40)$ \\
9  && $0.161\,587\,539\,919\,0\:(11)$ &&&
32 && $0.133\,030\,008\,805\:(14)$ &&&
55 && $0.073\,302\,598\,760\:(41)$ \\
10 && $0.161\,007\,527\,440\,5\:(14)$ &&&
33 && $0.131\,077\,399\,917\:(15)$ &&&
56 && $0.069\,959\,685\,549\:(43)$ \\
11 && $0.160\,366\,695\,552\,3\:(17)$ &&&
34 && $0.129\,066\,585\,774\:(16)$ &&&
57 && $0.066\,541\,237\,770\:(45)$ \\
12 && $0.159\,665\,113\,695\,7\:(20)$ &&&
35 && $0.126\,997\,670\,418\:(17)$ &&&
58 && $0.063\,042\,394\,713\:(46)$ \\
13 && $0.158\,902\,857\,464\,6\:(23)$ &&&
36 && $0.124\,870\,746\,321\:(18)$ &&&
59 && $0.059\,456\,825\,787\:(48)$ \\
14 && $0.158\,080\,008\,465\,4\:(27)$ &&&
37 && $0.122\,685\,891\,688\:(19)$ &&&
60 && $0.055\,776\,141\,692\:(51)$ \\
15 && $0.157\,196\,654\,156\,8\:(31)$ &&&
38 && $0.120\,443\,167\,320\:(20)$ &&&
61 && $0.051\,988\,975\,236\:(53)$ \\
16 && $0.156\,252\,887\,668\,9\:(35)$ &&&
39 && $0.118\,142\,612\,955\:(21)$ &&&
62 && $0.048\,079\,475\,600\:(56)$ \\
17 && $0.155\,248\,807\,597\,6\:(40)$ &&&
40 && $0.115\,784\,242\,975\:(22)$ &&&
63 && $0.044\,024\,687\,441\:(59)$ \\
18 && $0.154\,184\,517\,773\,3\:(45)$ &&&
41 && $0.113\,368\,041\,358\:(23)$ &&&
64 && $0.039\,789\,613\,916\:(63)$ \\
19 && $0.153\,060\,126\,999\,5\:(50)$ &&&
42 && $0.110\,893\,955\,708\:(24)$ &&&
65 && $0.035\,316\,860\,204\:(68)$ \\
20 && $0.151\,875\,748\,757\,7\:(55)$ &&&
43 && $0.108\,361\,890\,163\:(25)$ &&&
66 && $0.030\,501\,216\,714\:(75)$ \\
21 && $0.150\,631\,500\,875\,0\:(61)$ &&&
44 && $0.105\,771\,696\,922\:(26)$ &&&
67 && $0.025\,108\,988\,013\:(88)$ \\
22 && $0.149\,327\,505\,149\,1\:(66)$ &&&
45 && $0.103\,123\,166\,068\:(27)$ &&&
68 && $0.018\,331\,850\,81\:(13)$ \\ 
23 && $0.147\,963\,886\,924\,7\:(72)$ &&&
46 && $0.100\,416\,013\,257\:(28)$ \\
\hline
\end{tabular}
\end{center}
\end{table}
\end{landscape}
\end{document}